\documentclass[aps,prd,twocolumn,showpacs,superscriptaddress,linenumbers]{revtex4}
\usepackage{graphicx}
\usepackage{dcolumn}
\usepackage{bm}
\usepackage{amsmath}
\usepackage{amssymb}

\input babarsym

\newcommand{\BABARPubYear}       {13}
\newcommand{\BABARPubNumber}     {007}
\newcommand{\SLACPubNumber} {15644}

\long\def\inst#1{\par\nobreak\kern 4pt\nobreak
    {\it #1}\par\vskip 10pt plus 3pt minus 3pt}

\begin{document}

\begin{flushleft}
\babar-PUB-\BABARPubYear/\BABARPubNumber \\
SLAC-PUB-\SLACPubNumber \\
\end{flushleft}

{\pagestyle{empty}

\title{Measurement of the {\boldmath $B^+ \to \omega \ell^+ \nu$} branching fraction with semileptonically tagged {\boldmath $B$} mesons}

%
\author{J.~P.~Lees}
\author{V.~Poireau}
\author{V.~Tisserand}
\affiliation{Laboratoire d'Annecy-le-Vieux de Physique des Particules (LAPP), Universit\'e de Savoie, CNRS/IN2P3,  F-74941 Annecy-Le-Vieux, France}
\author{E.~Grauges}
\affiliation{Universitat de Barcelona, Facultat de Fisica, Departament ECM, E-08028 Barcelona, Spain }
\author{A.~Palano$^{ab}$ }
\affiliation{INFN Sezione di Bari$^{a}$; Dipartimento di Fisica, Universit\`a di Bari$^{b}$, I-70126 Bari, Italy }
\author{G.~Eigen}
\author{B.~Stugu}
\affiliation{University of Bergen, Institute of Physics, N-5007 Bergen, Norway }
\author{D.~N.~Brown}
\author{L.~T.~Kerth}
\author{Yu.~G.~Kolomensky}
\author{M.~J.~Lee}
\author{G.~Lynch}
\affiliation{Lawrence Berkeley National Laboratory and University of California, Berkeley, California 94720, USA }
% ...
\collaboration{The \babar\ Collaboration}
\noaffiliation

\begin{abstract}
We report a measurement of the branching fraction of the exclusive charmless semileptonic decay $B^+ \to \omega \ell^+ \nu$, where $\ell$ is either an electron or a muon. We use samples of $B^+$ mesons tagged by a reconstructed charmed semileptonic decay of the other $B$ meson in the event. The measurement is based on a dataset of 426.1 \invfb\ of \epem\ collisions at a center-of-mass energy of 10.58 \gev\ recorded with the \babar\ detector at the \pep2\ asymmetric-energy \epem\ storage rings. We measure a branching fraction of ${\cal B}(B^+ \to \omega \ell^+ \nu)$ = (1.35 $\pm$ 0.21 $\pm$ 0.11)$\cdot 10^{-4}$, where the uncertainties are statistical and systematic, respectively. We also present measurements of the partial branching fractions in three bins of \qsq, the invariant-mass squared of the lepton-neutrino system, and we compare them to theoretical predictions of the form factors.
\end{abstract}


\maketitle 

\setcounter{footnote}{0}

Measurements of branching fractions of charmless semileptonic $B$ decays can be used to determine the magnitude of 
the Cabibbo-Kobayashi-Maskawa matrix~\cite{ref:Cabibbo,ref:KM} element $|V_{ub}|$ and can thus provide an important constraint on the Unitarity Triangle. These measurements can either be inclusive, {\it i.e.}\ integrated over all possible hadronic final states, or exclusive, {\it i.e.}\ restricted to a specific hadronic final state, which is explicitly reconstructed. Studies of exclusive decays allow for more stringent kinematic constraints and better background suppression than inclusive decays. However, the predictions for exclusive decay rates depend on calculations of hadronic form factors, and these are affected by theoretical uncertainties different from those involved in inclusive decays.

Currently, the most precise determination of $|V_{ub}|$ with exclusive decays, both experimentally and theoretically, is based on a measurement of $B \to \pi \ell \nu$ decays~\cite{babarvub}. It is important to study decays to other pseudoscalar or vector mesons, in order to perform further tests of theoretical calculations, and to improve the knowledge of the composition of charmless semileptonic decays. We present measurements of the branching fractions ${\cal B}(B^+ \to \omega \ell^+ \nu)$, where $\ell = e, \mu$ and charge-conjugate modes are included implicitly. The $\omega$ meson is reconstructed in its decay to three pions, which has a branching fraction of ${\cal B}(\omega \to \pi^+\pi^-\pi^0) = (89.2 \pm 0.7)\%$~\cite{pdg}. This decay chain has previously been studied by the Belle collaboration using neutrino reconstruction~\cite{ref:belleomegalnu} and hadronic tagging~\cite{ref:belleomegalnu2}, and by the \babar\ collaboration using neutrino reconstruction on a partial~\cite{ref:babaromegalnu} and the full dataset~\cite{ref:wulsinomegalnu}, as well as employing a different analysis strategy based on a loose neutrino reconstruction technique~\cite{ref:montreal}; in this analysis, we employ a semileptonic tag on the full \babar\ dataset.

The results presented in this analysis are based on data collected with the \babar\ detector at the \pep2\ asymmetric-energy \epem\ storage rings, operating at the SLAC National Accelerator Laboratory. At \pep2, 9.0~\gev\ electrons collide with 3.1 \gev\ positrons to yield a center-of-mass (CM) energy of $\sqrt{s}$ = 10.58 \gev, which corresponds to the mass of the \FourS\ resonance. The asymmetric energies result in a boost of the CM frame of $\beta\gamma \approx$ 0.56. We analyse the full \babar\ dataset collected at the \FourS\ resonance from 1999 to 2008, corresponding to an integrated luminosity of 426 \invfb~\cite{ref:lumi} and 467.8 million \BB\ pairs. In addition, 40~fb$^{-1}$ are recorded at a CM energy 40 MeV below the $\Upsilon(4S)$ resonance to study background from $e^+e^- \to f\bar{f}~(f=u,d,s,c,\tau)$ continuum events.
 
A detailed description of the \babar\ detector can be found elsewhere~\cite{ref:BabarNIM}. Charged-particle trajectories are measured in a five-layer double-sided silicon vertex tracker and a 40-layer drift chamber, both operating in the 1.5-T magnetic field of a superconducting solenoid. Charged-particle identification is achieved through measurements of the particle energy loss (d$E$/d$x$) in the tracking devices and the Cherenkov angle obtained by an internally reflecting ring-imaging Cherenkov detector. A CsI(Tl) electromagnetic calorimeter provides photon detection and electron identification. Muons are identified in the instrumented flux return of the magnet.

In order to validate the analysis, Monte Carlo (MC) techniques are used to simulate the production and decay of \BB\ and continuum events~\cite{EvtGen,Jetset}, and to simulate the response of the detector~\cite{Geant}.
Charmless semileptonic decays are simulated as a mixture of three-body decays $B\to X_u \ell\nu \; (X_u = \pi, \eta, \eta', \rho, \omega)$ and are re-weighted according to the latest form-factor calculations from light-cone sum rules~\cite{BZ1,BZ2,BZ3}. Decays to non-resonant hadronic states $X_u$ with masses $m_{X_u} > 2m_{\pi}$ are simulated with a smooth $m_{X_u}$ spectrum~\cite{Defazio}.

Event-shape variables that are sensitive to the topological differences between jet-like continuum events and more spherical \BB\ events are used to suppress back- grounds from $e^+e^- \to q\bar{q}$ and other QED processes. We reject events for which the ratio of the second and zeroth Fox-Wolfram moments \cite{ref:Fox-Wolfram} is greater than 0.7. In addition, the event must contain at least six charged tracks (with three of them needed for the B tagging, as explained in the following), two of which must be identified as leptons of opposite charges.

In contrast to the earlier \babar\ $B^+ \to \omega \ell^+\nu$ measurements~\cite{ref:babaromegalnu,ref:wulsinomegalnu}, for the present analysis the second $B$ meson in the event is partially reconstructed and used as a tag $B$ that identifies the charge of the signal $B$ meson; this yields a smaller candidate sample, but with higher purity and  better signal discrimination.
We tag $B$ mesons decaying as $B^- \to D^{(*)}(X)\ell^-\bar{\nu}$ through the full hadronic reconstruction of \Dz\ mesons, where $(X)$ represents zero, one, or several pions in the final state, which are not explicitly reconstructed. The \Dz\ mesons are reconstructed via decays into \Km\pip, \Km\pip\piz, and \Km\pip\pip\pim. Neutral pions are reconstructed as $\piz \to \gaga$ with the requirement $115 < m_{\gaga} <$ 150 \mevcc\ on the diphoton invariant mass. Masses of $D$ candidates are required to be within 20 \mevcc\ of the nominal \Dz\ mass for $\Dz \to \Km\pip$ and $\Dz \to \Km\pip\pip\pim$ decays, and within 30 \mevcc\ for $\Dz \to \Km\pip\piz$ decays. We require the charged tracks from the \Dz\ decay to originate from a common vertex. We reconstruct \Dstarz\ mesons as $D^0\pi^0$. The mass difference between the \Dstarz\ candidate and its corresponding \Dz\ must be within 5 \mevcc\ of its expected value. Candidate $D^{(*)}$ mesons are paired with a charged lepton with absolute momentum (in the CM frame, denoted by a *) $|\vec{p}_{\ell}^{\,\ast}| > 0.8$ \gevc\ to form a $Y=D^{(\ast)}\ell$ candidate. The charged lepton is identified as either an electron or muon. The lepton identification efficiency is constant and greater than 92\% for electrons with momenta greater than 0.8 \gevc, and greater than 75\% for muons with momenta greater than 1.2 \gevc. The pion and kaon misidentification rates are of the order of 0.1\% and 0.5\%, respectively, for the electron selector, while both are below 5\% for the muon selector. The electron energy is corrected for bremsstrahlung photons emitted and detected close to the electron direction. The lepton and the kaon from the $D$ decay must have the same charge. Assuming that the $B^- \to Y \bar{\nu}$ decay hypothesis is correct, the angle $\theta_{BY}$ between the directions of the (measured) $Y$ and its parent $B$ is given by

\begin{equation}
\cos \theta_{BY} = \frac{2E_B E_Y -m_B^2 -m_Y^2}{2|\vec{p}_B||\vec{p}_Y|},
\label{eq:cosBY}
\end{equation}

\noindent where $E_B$, $m_B$, and $|\vec{p}_B|$ are the energy, mass, and absolute momentum of the $B$ meson, and $E_Y$, $m_Y$, and $|\vec{p}_Y|$ are the corresponding quantities for the $Y$ system. Eq. (\ref{eq:cosBY}) is valid in any frame of reference. In the CM frame however, the energy and momentum of the $B$ meson can be inferred from the beam energies, and $\cos \theta_{BY}$ can be calculated without any specific knowledge of the $B$ meson kinematics. If the $B^- \to Y \bar{\nu}$ hypothesis is correct, then $|\cos \theta_{BY}| \le 1$,  up to experimental resolution. Because $\cos \theta_{BY}$ is strongly correlated with our discriminating variable $\cos^2 \phi_B$ (described below), we impose the loose requirement $|\cos \theta_{BY}|\le 5$. The $B^- \to D^{(*)}(X)\ell^-\bar{\nu}$ tag efficiency is found to be 4.4\%.

After identifying the tag $B$, we require the remaining particles in the event to be consistent with a $B^+ \to \omega \ell^+ \nu$ decay, {\it i.e.}\ there should be exactly three additional tracks, one of them being identified as a charged lepton. We require the additional lepton to have an absolute CM momentum $|\vec{p}_{\ell}^{\,\ast}| > 0.8$ \gevc. The two remaining tracks (assumed to be pions and required to come from a common vertex) are combined with one neutral pion to form an $\omega$ candidate, which is required to have an invariant mass between 0.75 and 0.81 \gevcc. This is carried out with all the neutral pions in the event, because at this point we still allow for multiple $\omega$ candidates in each event. These $\omega$ candidates are then paired with the lepton to form $X=\omega\ell$ candidates. The angle $\theta_{BX}$ is defined analogously to $\theta_{BY}$ in Eq. (\ref{eq:cosBY}), and we require $|\cos\theta_{BX}| \le 5$. Since the signal decay is charmless, the momenta of the daughter particles tend to be relatively large; we thus require $|\vec{p}_{\omega}^{\,\ast}| + |\vec{p}_{\ell}^{\,\ast}| > 2.5$ \gevc, where $|\vec{p}_{\omega}^{\,\ast}|$ is the absolute CM momentum of the $\omega$ candidate.

We also reject events containing lepton pairs kinematically and geometrically consistent with having originated from the decay of a \jpsi\ meson. If the two leptons are an \epem\ pair, we require them to be inconsistent with $\g \to \epem$ conversion.

For each combination of $D^{(*)}\ell^-$ and $\omega \ell^+$ candidates, we require that there be no additional tracks in the event and less than 200 MeV of energy from photon candidates not associated with the reconstructed event. In the case that more than one $D^{(*)}\ell^- - \omega \ell^+$ combination satisfies all requirements, which is the case in 76.1\% of the events, a single candidate is chosen by the following method: if a $Y=D^{\ast}\ell^-$ is reconstructed, all $Y=D\ell^-$ candidates reconstructed with the same $D$ are discarded. Among the remaining multiple $Y=D^{\ast}\ell^-$ or $Y=D\ell^-$ candidates, those with the reconstructed $D$ mass closest to the nominal value are selected. If several candidates fall into this category ({\it i.e.}\ events with multiple $X=\omega\ell$ candidates), we select the candidate with the smallest absolute value of $\cos\theta_{BY}$ and $\cos\theta_{BX}$, in that order.

The momentum vectors of the reconstructed $Y$ and $X$ systems together define a plane. The angles between the momentum vectors of $Y$ and $X$ relative to the momentum of the corresponding parent $B$ meson, $\theta_{BY}$ and $\theta_{BX}$, are calculated in the CM frame using the known beam energies, so that the possible $B$ directions are constrained to two cones around $\vec{p}_{Y(X)}^{\,\ast}$ with angles $\theta_{BY (BX)}$, respectively. This information, together with the requirement that tag and signal $B$ mesons emerge back-to-back in the CM frame, determines the direction of either $B$ meson up to a two-fold ambiguity. A schematic of the event kinematics is shown in Fig.~\ref{fig:phib}. The angle between the $Y-X$ plane and either $\vec{p}_B^{\,\ast}$ possibility, denoted by $\phi_B$, is given by

\begin{equation}
\cos^2 \phi_B = \frac{\cos^2 \theta_{BY}+2\cos\gamma\cos\theta_{BY}\cos\theta_{BX}+\cos^2\theta_{BX}}{\sin^2\gamma},
\end{equation}

\noindent where $\gamma$ is the angle between the $X$ and $Y$ momenta in the CM frame. Events consistent with the hypothesis of two semileptonic $B \to Y (X) \nu$ decays have $\cos^2 \phi_B \le$ 1, up to resolution effects. 

\begin{figure}
\begin{center}
\includegraphics[width=0.495\textwidth,totalheight=0.24\textheight]{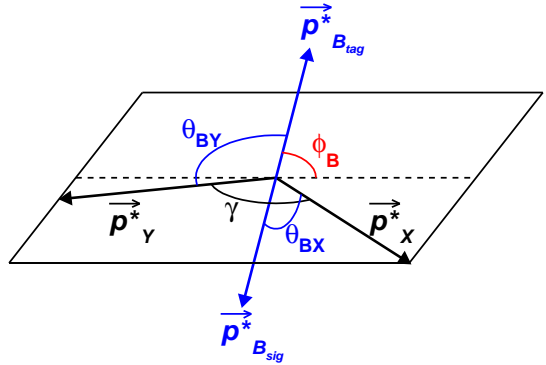}
\caption{Event kinematics of a decay $B_{\rm sig} \to X \nu$, $B_{\rm tag} \to Y \nu$. The various angles and momenta are described in the text.}
\label{fig:phib}
\end{center}
\end{figure}

We extract the signal yield from an extended binned maximum-likelihood fit to the measured $\cos^2\phi_B$ distribution. The data are described as a sum of three contributions, $dN/d \cos^2 \phi_B = N_{\rm sig}{\cal P}_{\rm sig} + N_{\rm bkg}{\cal P}_{\rm bkg} + N_{\rm cont}{\cal P}_{\rm cont}$, where the $N_{i}$ and ${\cal P}_{i}$ are the yields and probability density functions (PDFs) of: signal (``sig''), $B\bar{B}$ background (``bkg''), and background from continuum events (``cont''). The signal PDF, ${\cal P}_{\rm sig}$, is parametrized as a threshold function in the physical region ($0 \le \cos^2\phi_B \le 1$) with finite resolution and an exponential tail:

\begin{equation}\label{sigPDF}
\mathcal{P}_{\rm sig} \propto \frac{1-\mathrm{erf}[p_0\log(p_1 \cos^2\phi_B)]}{2} + p_2\mathrm{e}^{-p_3 \cos^2\phi_B}.
\end{equation}

\noindent The \BB\ background and continuum background PDFs are both modelled as the sum of an exponential function and a positive constant:
\begin{equation}\label{bkgPDF}
\mathcal{P}_{\rm bkg} \propto \mathrm{e}^{-p_4 \cos^2\phi_B} + p_5^2,
\end{equation}
\begin{equation}
\mathcal{P}_{\rm cont} \propto \mathrm{e}^{-p_6 \cos^2\phi_B} + p_7^2.
\end{equation}

\noindent The various yields are obtained from a binned maximum likelihood fit (see Fig.~\ref{cos2PhiB}) of $dN/d \cos^2 \phi_B$ to the data, where the PDF shape parameters of the three contributions are fixed to those values obtained from three separate fits to the corresponding MC samples. The yield of the continuum contribution however is fixed to the luminosity-adjusted value from the MC sample, instead of being allowed to float, due to its similar functional form as the background PDF, and its small overall size.
We find 103 $\pm$ 16 signal events and 355 $\pm$ 23 background events. The dominant contribution to background events comes from $B^+ \to X_c \ell^+ \nu$ events, with most of the $B^+ \to X_u \ell^+ \nu$, other \BB, and $q\bar{q}$ backgrounds eliminated at the end of the event and final candidate selection. The $B^+ \to \omega \ell^+ \nu$ signal efficiency is 2.4\% after all selection cuts.

The branching fraction is given by

\begin{equation}\label{BF}
{\cal B} (B^+ \to \omega \ell^+ \nu) = \frac{(N_{\rm sig}/ \varepsilon) \cdot r_{\varepsilon}^{\rm tag}}{4 \cdot N_{\BpBm} \cdot {\cal B}(\omega \to \pip\pim\piz)},
\end{equation}
\noindent where $N_{\rm sig}$ is the number of reconstructed signal events, $\varepsilon$ is the reconstruction efficiency, and $N_{\BpBm}$ is the number of produced \BpBm\ events, which is given by $N_{\BpBm} = f_{+-}/f_{00} \cdot N_{\BB}/2$, where $f_{+-}/f_{00}$ is the ratio of the \upsbpbm\ and \upsbzbz\ branching fractions~\cite{ref:HFAG}. The factor of 4 arises as the product of two contributions: one factor of 2 comes from the fact that the branching fraction is quoted as the average of the electron and muon contributions, and another factor of 2 from the fact that either of the two $B$ mesons in the \BpBm\ event may decay in the signal mode. The tag efficiency correction factor $r_{\varepsilon}^{\rm tag}$ takes into account differences in the tagging efficiency between data and simulation, including all tag side branching fractions and reconstruction efficiencies, and is determined by studying ``double tag'' events, {\it i.e.}\ events reconstructed as \BB\ with both $B$ mesons decaying as $B \to D^{(*)}\ell \nu$.

\begin{figure}[!h]
\begin{center}
\includegraphics[width=0.51\textwidth,totalheight=0.3\textheight]{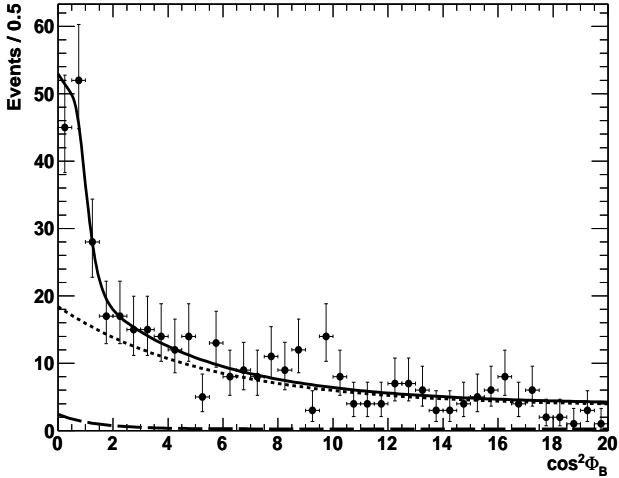}
\caption{Distribution of $\cos^2\phi_B$. Points represent data, the curves are stacked fit results for ``continuum background'' (dashed), ``\BB\ background'' (dotted), and ``signal'' (solid) PDFs, as defined in the text.}\label{cos2PhiB}
\end{center}
\end{figure}

We also measure the partial branching fraction $\Delta{\cal B}/ \Delta q^2$ in bins of \qsq, the invariant-mass squared of the lepton-neutrino system. For a $B^+ \to \omega \ell^+ \nu$ decay, \qsq\ is calculated in the approximation that the $B$ is at rest in the CM frame, {\it i.e.}\ $\qsq = (m_B - E_{\omega}^{\,\ast})^2 - |\vec{p}_{\omega}^{\,\ast}|^2$, where $E_{\omega}^{\,\ast}$ and $|\vec{p}_{\omega}^{\,\ast}|$ are the energy and absolute momentum of the $\omega$ meson in the CM frame. We divide the data  into three bins: $q^2 < 7, 7 \le q^2 < 14$ and $q^2 \ge 14$ GeV$^2/c^4$, in each of which the yield is extracted separately using the same maximum likelihood fit as for the full branching fraction. The \qsq\ resolution is 0.2 GeV$^2/c^4$, significantly smaller than the widths of the \qsq\ bins used to measure the partial branching fractions. Table~\ref{tab:q2} summarizes the measured partial branching fractions for these three regions of \qsq\, along with the corresponding signal yields and overall reconstruction efficiencies (including the fit), which are determined from MC signal events. The MC simulation is validated by detailed comparisons with data at various stages in the selection process, and the corresponding uncertainties are taken into account in the systematic error analysis, as discussed in the following. In Fig.~\ref{q2}, the measured partial branching fractions are compared to the predicted $\qsq$ dependence by Ball-Zwicky~\cite{BZ1,BZ2,BZ3} and ISGW2~\cite{ref:ISGW2} calculations, normalized to the measured total branching fraction. Within the large experimental uncertainties, both form factor calculations are consistent with the data.

\begin{figure}
\begin{center}
\includegraphics[width=0.51\textwidth,totalheight=0.3\textheight]{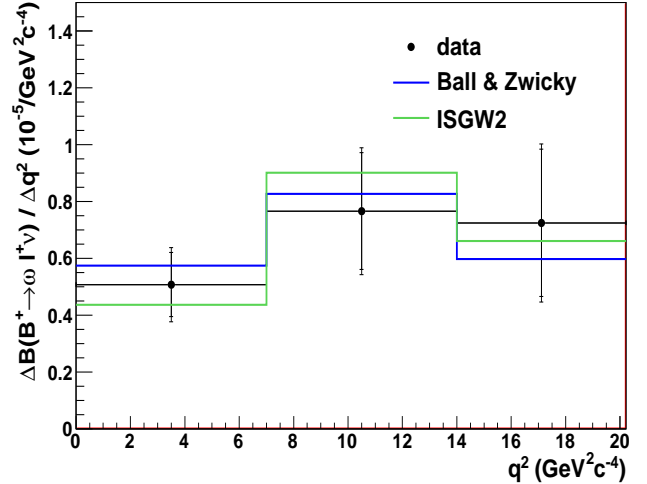}
\caption{The partial branching fractions $\Delta {\cal B} (B^+ \to \omega \ell^+\nu) / \Delta \qsq$ in bins of \qsq. The points are the measurements, with the error bars indicating the statistical uncertainty, and the quadratic sum of statistical and systematic uncertainties, respectively. The histograms are predictions by Ball-Zwicky (blue), and ISGW2 (green) calculations, each scaled to the measured total branching fraction.}
\label{q2}
\end{center}
\end{figure}

\begin{table}
\caption{Signal yields, reconstruction efficiencies, and partial branching fractions $\Delta {\cal B} (B^+ \to \omega \ell^+\nu)$ in three bins of $\qsq$, along with the corresponding values over the full range. The uncertainties are statistical and systematic, respectively.}
\label{tab:q2}
\begin{center}
\begin{tabular}{r@{\;\qsq}l|r@{$\;\pm\;$}l|c|r@{$\;\pm\;$}l}
\hline\hline
\multicolumn{2}{c|}{\qsq\ bin} & \multicolumn{2}{c|}{signal yield} & efficiency & \multicolumn{2}{c}{$\Delta {\cal B}(B^+ \to \omega \ell^+\nu)$}\\
\multicolumn{2}{c|}{(GeV$^{2}/c^{4}$)} & \multicolumn{2}{|c|}{} & ($10^{-4}$) & \multicolumn{2}{c}{$(10^{-5})$}\\
\hline
& $<$ 7 & 35 & 8  & 10.6 & 3.6 & 0.8 $\pm$ 0.5\\
7 $\le$ & $<$ 14 & 39 & 10 & 7.91 & 5.4 & 1.4 $\pm$ 0.6\\
14 $\le$ & & 28 & 9 & 6.81 & 4.5 & 1.5 $\pm$ 0.6\\
\hline
\multicolumn{2}{c|}{Total} & $\;\;$103 & 16 & 8.41 & $\;$13.5 & 2.1 $\pm$ 1.1\\
\hline
\end{tabular}
\end{center}
\end{table}

\begin{table}
\caption{Summary of the relative systematic uncertainties for the total branching fraction $\mathcal{B}(B^+ \to \omega \ell^+ \nu)$.}
\label{tab:Systematics}
\begin{center}
\begin{tabular}{l|c}
\hline\hline
Source of uncertainty & $\delta \mathcal{B}(B^+ \to \omega \ell^+\nu)$ (\%)\\
\hline
Additive errors &\\
\hline
\hspace{0.2cm} Fit yield & 3.9\\
\hspace{0.2cm} Fit bias & 0.3\\
\hline
Multiplicative errors &\\
\hline
\hspace{0.2cm} Tracking efficiency & 1.6\\
\hspace{0.2cm} PID efficiency & 3.0\\
\hspace{0.2cm} $\piz$ efficiency & 3.4\\
\hspace{0.2cm} $B \to \omega \ell \nu$ FF & 1.1\\
\hspace{0.2cm} $\mathcal{B} (\omega \to \pi^+\pi^-\pi^0)$ & 0.8\\
\hspace{0.2cm} $B \to D^{(\ast)} \ell \nu$ FF & 1.1\\
\hspace{0.2cm} $\mathcal{B}(B \to D^{(\ast,\ast\ast)} \ell \nu)$ & 2.0\\
\hspace{0.2cm} $B$ counting & 1.1\\
\hspace{0.2cm} $f_{+-}/f_{00}$ & 2.7\\
\hspace{0.2cm} Tag efficiency & 3.2\\
\hspace{0.2cm} MC statistics & 2.1\\
\hline
Total systematic error & 8.3\\
\hline\hline
\end{tabular}
\end{center}
\end{table}

The systematic uncertainties on the measured branching fraction are listed in Table~\ref{tab:Systematics}. They are estimated by varying the detection efficiencies or the parameters that impact the modelling of the signal and the background processes within their uncertainties. The complete analysis or only parts of it are then repeated and the differences in the resulting branching fractions are taken as the systematic errors. The total systematic error is obtained by adding in quadrature all listed contributions.

To estimate the uncertainty related to the stability of the yield extraction fit, we vary each parameter of the fit individually within its uncertainty derived from MC statistics, and also the functional forms of the PDFs used for the yield extraction; we find a maximum deviation of four events from varying the background parameters, corresponding to a fit yield uncertainty of 3.9\%. To estimate the uncertainty due to a potential fit bias, we randomly fluctuate the individual signal, background, and continuum yields about their expected values according to Poisson statistics, and generate toy MC samples from the sum of these contributions. A fit is then applied in the usual way, and the deviation of the mean of the obtained pull distribution for the signal yield from the expected value of zero is quoted as the fit bias uncertainty of 0.3\%.

Uncertainties due to the reconstruction of charged particles are evaluated by varying their corresponding reconstruction efficiencies in the simulation, and comparing the resulting efficiencies to the original ones. As double tag events are used to determine the $D^{(*)} \ell \nu$ reconstruction efficiency, detector simulation uncertainties are applied only to particles on the signal side: 0.5\% per track and 3.4\% per \piz.  For lepton identification, relative uncertainties of 1.4\% and 3\% are used for electrons and muons, respectively. The tag efficiency uncertainty of 3.2\% is derived from the limited statistics of the double tag sample and from the difference in tagging efficiency found between double tag and single tag samples, added in quadrature.

Uncertainties in the modelling of the signal and tag decays due to the imperfect knowledge of the form factors affect the shapes of kinematic spectra and thus the acceptances of signal events. We use the Isgur-Wise quark model \cite{ref:ISGW2} as an alternative to the default Ball-Zwicky calculations~\cite{BZ1,BZ2,BZ3} to test the model dependence of the $B^+ \to \omega \ell^+ \nu$ simulation. The uncertainties due to the imperfect MC modeling on the tag side are similarly evaluated by reweighting the $B^- \to D^{(*)} \ell^- \bar{\nu}$ form factors, and also by varying the $B^- \to D^{(*,**)} \ell^- \bar{\nu}$ branching fractions. We also include a 1.1\% systematic uncertainty from counting \BB\ pairs~\cite{ref:Bcounting}, a 0.8\% systematic uncertainty from the $\omega \to \pip\pim\piz$ branching fraction~\cite{pdg}, and a 2.7\% systematic uncertainty from the correction factor $f_{+-}/f_{00} = 1.056 \pm 0.028$~\cite{ref:HFAG}.

In summary, we have measured the total branching fraction of the charmless semileptonic decay $B^+ \to \omega \ell^+ \nu$ to be
\begin{equation}
\cal{B} (B^+ \to \omega \ell^+ \nu) = ({\rm 1.35} \pm {\rm 0.21} \pm {\rm 0.11}) \cdot {\rm 10^{-4}}
\end{equation}

\noindent where the errors are statistical (data and simulation) and systematic, respectively.
This result is consistent with the current world average~\cite{pdg} and previous \babar\ results~\cite{ref:babaromegalnu,ref:wulsinomegalnu,ref:montreal}, and manifests a slight improvement over the earlier measurements from Belle~\cite{ref:belleomegalnu,ref:belleomegalnu2}.

The value of $|V_{ub}|$ can be determined from the measured partial branching fraction, the \Bp\ lifetime $\tau_+$, and the reduced partial decay rate $\Delta\zeta$ of the corresponding theoretical form factor model:

\begin{equation}
|V_{ub}| = \sqrt{\frac{\Delta\mathcal{B}(q^2_{\rm min},q^2_{\rm max})}{\tau_+ \Delta\zeta(q^2_{\rm min},q^2_{\rm max})}}
\end{equation}
\begin{equation}
\Delta\zeta(q^2_{\rm min},q^2_{\rm max}) = \frac{1}{|V_{ub}|^2} \int_{q^2_{\rm min}}^{q^2_{\rm max}}\frac{{\rm d}\Gamma_{\rm theory}}{{\rm d}q^2} {\rm d}q^2
\end{equation}

Form-factor calculations are available from the method of light-cone sum rules (LCSR)~\cite{BZ2} and the ISGW2 quark model~\cite{ref:ISGW2}. With $\Delta\zeta = 7.10 \, (7.02) \ps^{-1}$ for the LCSR (ISGW2) model, and $\tau_+ = (1.638 \pm 0.011)$ ps~\cite{pdg}, we obtain
\begin{equation}
\Vub\ = \begin{cases} (3.41 \pm 0.31)\cdot 10^{-3} & \text{for LCSR}\\
(3.43 \pm 0.31)\cdot 10^{-3} & \text{for ISGW2} \end{cases}
\end{equation}

\noindent where the quoted uncertainty does not include any uncertainty from theory, since uncertainty estimates of the form-factor integrals are not available. Both form-factor calculations arrive at very similar values for $|V_{ub}|$, which are consistent with the values derived from other exclusive semileptonic $B$ decays~\cite{ref:Vera,ref:wulsinomegalnu}.

We combine the measurement presented here with the combination of the later two~\cite{ref:wulsinomegalnu,ref:montreal} of the three previous untagged \babar\ measurements that is presented in Ref.~\cite{ref:montreal}. The measurements are combined using the BLUE (Best Linear Unbiased Estimate) technique~\cite{ref:Lyons}, where the correlation of the statistical uncertainties between this analysis and the combination of the two untagged \babar\ analyses is negligible (7\%). The correlation of the systematic uncertainties between this measurement and the combination of the two untagged \babar\ measurements is estimated to be 74\%, based on the systematic uncertainty contributions which a given pair of analyses has in common. The combined average of the three measurements is $\mathcal{B}(\Bu \to \omega \ellp \nu) = (1.23 \pm 0.10 \pm 0.09)\cdot 10^{-4}$.

We are grateful for the excellent luminosity and machine conditions
provided by our \pep2\ colleagues, 
and for the substantial dedicated effort from
the computing organizations that support \babar.
The collaborating institutions wish to thank 
SLAC for its support and kind hospitality. 
This work is supported by
DOE
and NSF (USA),
NSERC (Canada),
CEA and
CNRS-IN2P3
(France),
BMBF and DFG
(Germany),
INFN (Italy),
FOM (The Netherlands),
NFR (Norway),
MES (Russia),
MINECO (Spain),
STFC (United Kingdom). 
Individuals have received support from the
Marie Curie EIF (European Union)
and the A.~P.~Sloan Foundation (USA).

\end{document}